\documentclass[final]{appolb}
\usepackage{epsfig}
\usepackage{amsmath}

\begin{document}
\title{Saturation and Vector Mesons%
\thanks{Work supported by 
QCDNET contract FMRX-CT98-0194.}
}
\author{St\'ephane Munier
\address{Universit\`a di Firenze,
via Sansone 1 \\
50009 Sesto Fiorentino, Italy} 
}
\maketitle
\begin{abstract}
Through an analysis of
diffractive vector meson production,
we show that the proton appears quite dense
to a small size probe at present HERA energies.
This means that saturation effects are already
important.
\end{abstract}

\section{Introduction}

The question whether 
traditional leading twist QCD is the relevant
description of HERA small-$x$ data has still no clear-cut
answer. A particularly successful saturation model proposed
by Golec-Biernat and W\"usthoff (GBW) \cite{GBW} suggests that 
the common picture of a proton looking 
like a dilute set of partons might be unjustified.
We have recently shown {\em directly} that
the proton is a dense system already at present HERA
energies \cite{ref}.

The method we used is summarized
in the next section.
We then dedicate a section to a by-product of our analysis:
the measurement of total dipole-proton cross section.

\section{A picture of the proton at HERA}

Wave diffraction allows to obtain a picture of a
microscopic object. 
A Fourier transform relates 
its density profile to the square root of the intensity 
of light measured on the interference pattern.

It turns out that one can analyse the HERA data for
diffractive vector meson production in this way.
Indeed, at high energy, such a process is equivalent to
{elastic} diffusion of dipoles. 
The picture is the following \cite{DIPOLENIK}: 
in an appropriate reference frame,
the photon 
of virtuality $Q^2$ splits in a $q\bar q$ dipole
which scatters elastically off the target proton before
recombining into a meson.
The size of the interacting dipole is distributed 
around a mean value $r_Q$
and the total flux $N(Q)$ is the scalar product of
photon and meson wave functions 
$\langle\psi_{\gamma^*}|\psi_V\rangle$.
The density profile is extracted by taking the Fourier transform
of the square root of the differential cross section 
$d\sigma/dt$ with respect
to the momentum transfer $\Delta$ ($t=-\Delta^2$).
The variable conjugated to $\Delta$ is the impact parameter
of the dipole relatively to the center of the proton.
One then sees that this profile is 
related to the $S$-matrix element for elastic
dipole-proton scattering
(a key observation is that $S$ is essentially real
at high energy). It reads
\begin{equation}
S(x,r_Q,b)=1-\frac{1}{2\pi^{3/2}N(Q)}\int d^2\Delta e^{-i\Delta b}
\sqrt{\frac{d\sigma}{dt}}\ .
\label{eq:s}
\end{equation}
As a phenomenological model is needed for the meson
wave function $|\psi_V\rangle$ which appears in $N(Q)$, 
checks of model-independence of
the method are in order.
We checked that the dipole size is indeed distributed around
a mean size $r_Q\!=\!A/\sqrt{Q^2\!+\!M_V^2}$ independently of the model
($M_V$ is the mass of the meson).
We also verified that the dipole flux $N(Q)$ is strongly constrained
by the well-known photon wave function $|\psi_{\gamma^*}\rangle$, and thus 
is quite model-independent.

We applied formula~(\ref{eq:s}) 
to the ZEUS data for diffractive electroproduction of 
longitudinal $\rho$ mesons \cite{zeus}. 
As saturation is a high energy effect, we considered the
lowest available values of $x$ ($\sim\!5\cdot 10^{-4}$) only.
We varied the virtuality $Q^2$ of the photon so that the
effective scale $Q^2\!+\!M_V^2$ is always larger than 
$1\ \mbox{GeV}^2$.
The result of our analysis for 3 different values
of the photon virtuality is represented on Fig.1.
The estimated error shown stems from the lack of data for
large momentum transfer $t\!>\!0.6\ \mbox{GeV}^2$.
It does not include experimental errors on the measured
quantities.

Why is $S$ a good estimator of the importance of
saturation effects?
As argued before, the value of $S$ tells us about how dense
the proton looks. $S\!=\!0$ means blackness: it
is the unitarity limit.
For a more quantitative interpretation,
one observes that $1\!-\!S^2$ is the interaction
probability of a dipole that hits the proton at impact parameter $b$.
If this probability is significant, it means that
more rescatterings are likely to occur. 
We see that the interaction probability 
is more than 50\% and
could already
reach 75\% at the center of the proton.

A more common parameter characterizing the saturation regime
is the saturation scale $Q_s^2$. 
Roughly speaking, it is defined as the
maximal virtuality for which the photon sees the proton 
as a dense medium.
We found $Q_s^2$ of order 1 to 1.5~$\mbox{GeV}^2$ near the
center of the proton, and 0.2~$\mbox{GeV}^2$
on the periphery.

We refer the interested reader to Ref.\cite{ref} for all details
of the analysis and more results.

\begin{figure}[ht]
\begin{center}
\epsfig{file=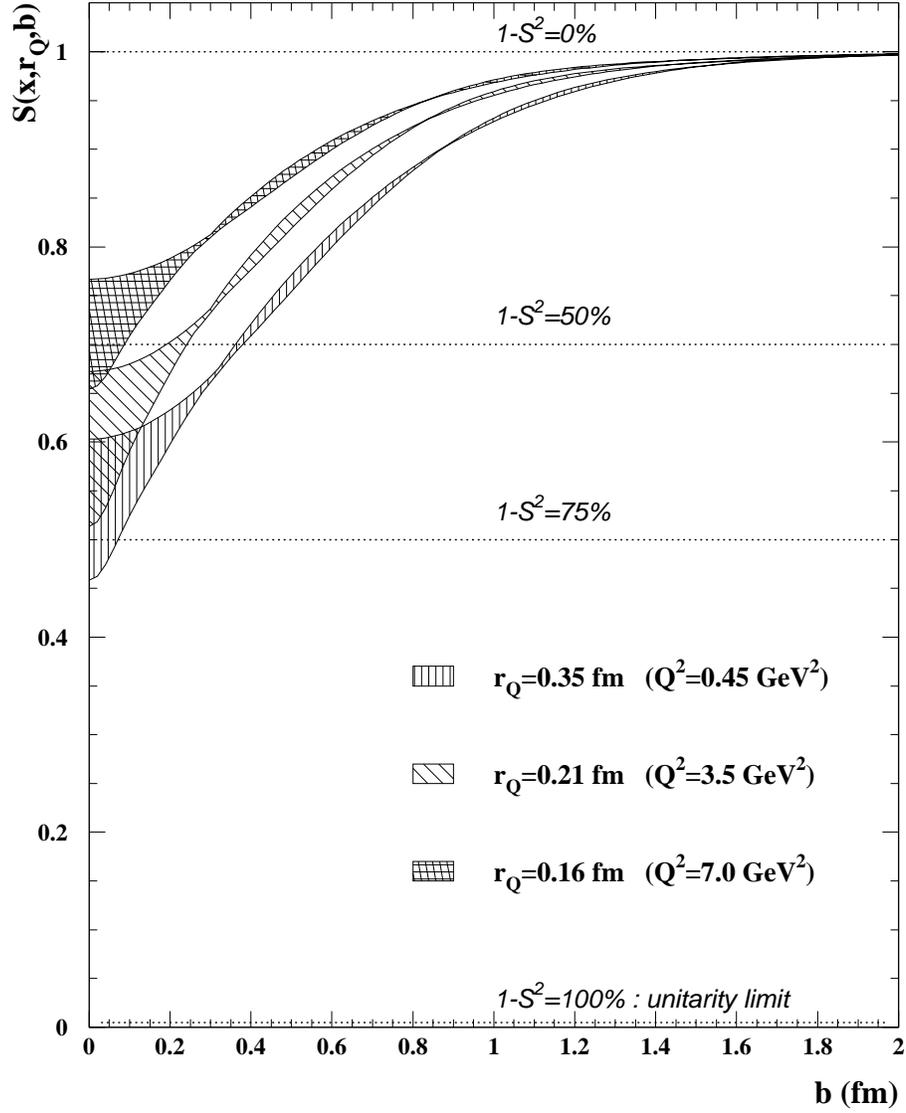,width=12.9cm}
\end{center}
\caption{$S$-matrix as a function of the impact parameter
for $x\!\sim\!5\cdot 10^{-4}$ and $Q^2=0.45,\ 3.5,\ 7\ \mbox{GeV}^2$.
The width of the bands represents the uncertainty due
to the lack of experimental data for $t>0.6\ \mbox{GeV}^2$. 
It is obtained by extrapolating the cross section
with functions of $t$ 
with behaviour between $t^{-3}$ and $e^{-\lambda t}$.
}
\end{figure}

\clearpage

\section{Extracting the dipole-proton total cross section}

\begin{figure}[ht]
\begin{center}
\epsfig{file=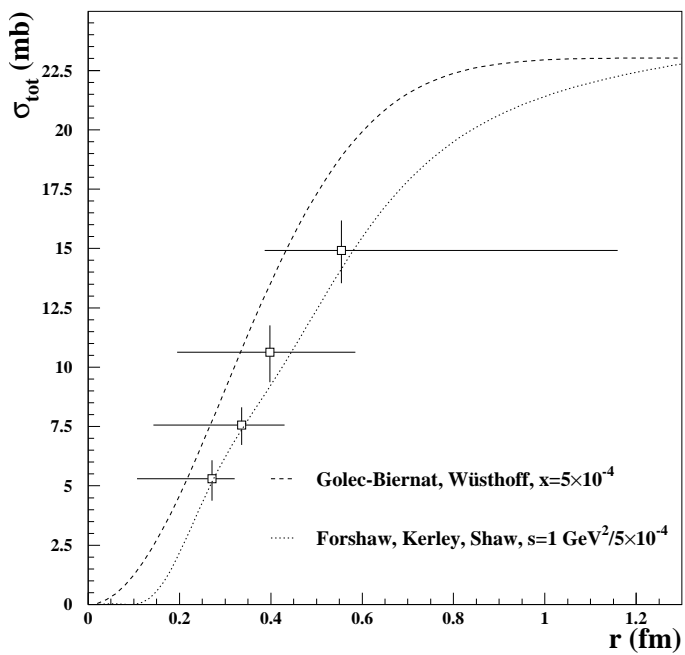,width=10.cm}
\end{center}
\caption{The total dipole-proton cross section for different values
of the dipole radius~$r$. 
The errors on the cross section are estimated from experimental errors.
The error bars on~$r$ are due to the width of the dipole distribution at
the $\gamma^*\!-\!\rho$ vertex.
The points represent the average radius obtained with a weight
given by GBW cross section.
For comparison,
in dashed line: Golec-Biernat and W\"usthof model~\cite{GBW};
dotted line: Forshaw, Kerley, Shaw model \cite{Forshaw}.}
\label{fig:stot}
\end{figure}

Having the $S$-matrix, one can 
easily obtain the total dipole-proton cross section.
It is an interesting quantity because of its universality.
One needs the forward differential cross section
from the data and the dipole flux
$N(Q)$ and average dipole size $r_Q$ from the model.
The cross section reads
\begin{equation}
\sigma_{tot}(x,r_Q)=\frac{4\sqrt{\pi}}{N(Q)}
\sqrt{\frac{d\sigma}{dt}}_{|t=0}\ .
\end{equation}
In pratice, we give an upper and a lower bound for $r_Q$.
An approximate lower bound is obtained when the 
dipole size distribution is
weighted by a flat (saturated) cross section 
(i.e. $\sigma_{tot}(r)\!=\!\mbox{cst}$). The upper bound
comes for  a colour transparent cross section 
$\sigma_{tot}(r)\!\propto\! r^2$.
A realistic average value can be computed by taking
an interpolating weight.
We took GBW cross section,
and we checked that the expectation value $r_Q$ is
not very much dependent on the exact
point where the transition occurs, as long as it is around
$r_s\!\sim\! 1\ \mbox{GeV}^{-1}$. The result is plotted
on Fig.\ref{fig:stot}.

\section{Summary}

We have shown that perturbative
saturation effects are already important at HERA.
Our argument was that the proton appears 
quite dense to a relatively small size probe.
We were able to give a quantitative estimate of
its darkness by measuring the $S$-matrix element for
elastic dipole-proton scattering. 
It turns out that the probability 
that a dipole of size $0.2$~fm undergoes an
inelastic interaction 
is more than $50\%$ at small impact
parameters, and for the high energy data
from HERA. This translates into a saturation scale $Q_s^2$
lying between $1$ and $1.5\ \mbox{GeV}^2$
which is consistent with the assumptions contained
in GBW model.
More exclusive data for quasi-elastic processes
at large momentum transfer on one hand, and 
impact parameter-dependent analysis on the other hand,
would be very welcome in the
future since saturation effects are expected to be maximal for
central collisions.\\

\noindent
[{\em 
The interesting question whether
this statement is in contradiction
with the successful 
description of vector meson photoproduction at large $t$
by 2-gluon exchange enhanced by BFKL corrections was raised by
Martin McDermott during the discussion.
From our result, for high energies ($W^2\sim 10^4\ \mbox{GeV}^2$),
low $Q^2$ and large $t$ diffractive processes, 
one would expect significant saturation effects.
In any case, it would be very instructive
to construct a model of GBW type for large $t$ processes.
}]




\begin{thebibliography}{999}

\bibitem{GBW}  K. Golec-Biernat, M. W\"usthoff,
             {\it Phys. Rev.} {\bf D59}  (1999) 014017.

\bibitem{ref}
S.~Munier, A.~M.~Sta\'sto, A.~H.~Mueller,
{\em Nucl.\ Phys.}\ {\bf B603} (2001) 427.


\bibitem{DIPOLENIK}
N.N. Nikolaev and B.G. Zakharov, {\em Z. Phys.} {\bf C49}
(1991) 607; {\em Z. Phys} {\bf C53} (1992) 331; {\em Z. Phys.} {\bf C64}
(1994) 651; {\em JETP} {\bf 78} (1994) 598;
J. Nemchik, N.N. Nikolaev, B.G. Zakharov,
 {\it Phys.Lett.} {\bf B341} (1994) 228; A.~H.~Mueller,
Nucl.\ Phys. {\bf B415} (1994) 373.

\bibitem{zeus}
J.~Breitweg {\it et al.}  [ZEUS Collaboration],
Eur.\ Phys.\ J.\ C {\bf 6} (1999) 603.


\bibitem{Forshaw}
J.~R.~Forshaw, G.~Kerley and G.~Shaw,
Phys.\ Rev.\ D {\bf 60} (1999) 074012.






\end{thebibliography}
\end{document}